\documentclass[fleqn,usenatbib,useAMS]{mnras}
\usepackage{graphicx}   
\usepackage{amsmath}    
\usepackage{amssymb}    
\usepackage{multicol}        
\usepackage{bm}         
\usepackage{pdflscape}  
\usepackage[T1]{fontenc}
\usepackage{ae,aecompl}
\usepackage{newtxtext,newtxmath}

\def\kms{km s$^{-1}$}

\def\aap{A\&A}
\def\apjl{ApJ}
\def\apj{ApJ}
\def\apjs{ApJS}
\def\aj{AJ}
\def\mnras{MNRAS}
\def\nat{Nature}
\def\pasp{PASP}

\title[Double-lined WDs]{Hidden in Plain Sight: A Double-lined White Dwarf Binary 26 pc Away and a Distant Cousin}

\author[Kilic, B{\'e}dard, \& Bergeron]
{Mukremin Kilic$^1$, A. B{\'e}dard$^2$, P. Bergeron$^2$\\
$^1$Homer L. Dodge Department of Physics and Astronomy, University of Oklahoma, 440 W. Brooks St., Norman, OK, 73019, USA\\
$^2$D\'epartement de Physique, Universit\'e de Montr\'eal, C.P. 6128, Succ. Centre-Ville, Montr\'eal, QC H3C 3J7, Canada\\}

\date{\ \ Submitted \today \vspace{-0.5cm}}
\pubyear{2020}

\begin{document}
\label{firstpage}
\pagerange{\pageref{firstpage}--\pageref{lastpage}}
\maketitle

\begin{abstract}

We present high-resolution spectroscopy of two nearby white dwarfs with inconsistent spectroscopic and parallax distances.
The first one, PG 1632+177, is a 13th magnitude white dwarf only 25.6 pc away. Previous spectroscopic observations failed
to detect any radial velocity changes in this star. Here, we show that PG 1632+177 is a 2.05 d period double-lined
spectroscopic binary (SB2) containing a low-mass He-core white dwarf with a more-massive, likely CO-core white dwarf companion.
After L 870-2, PG 1632+177 becomes the second closest SB2 white dwarf currently known. Our second target,
WD 1534+503, is also an SB2 system with an orbital period of 0.71 d. For each system, we constrain the atmospheric
parameters of both components through a composite model-atmosphere analysis. We also present a new set of NLTE
synthetic spectra appropriate for modeling high-resolution observations of cool white dwarfs, and show that NLTE effects
in the core of the H$\alpha$ line increase with decreasing effective temperature. We discuss the orbital period and mass
distribution of SB2 and eclipsing double white dwarfs with orbital constraints, and demonstrate that the observed population
is consistent with the predicted period distribution from the binary population synthesis models. The latter predict more
massive CO + CO white dwarf binaries at short ($<1$ d) periods, as well as binaries with several day orbital periods;
such systems are still waiting to be discovered in large numbers.

\end{abstract}


\section{Introduction}

Double-lined spectroscopic binaries (SB2) are the best: radial velocity measurements of both stars in the system enable
a direct measurement of the gravitational redshifts, masses, the mass ratio, and the inclination of the binary. However, double-lined binaries are
hard to identify in low-resolution spectroscopy that is typical in large scale surveys like the Sloan Digital Sky Survey.
This is one of the challenges that prevents us from detecting the double white dwarf progenitors of type Ia supernovae
\citep{rebassa19}.

Population synthesis models indicate that double white dwarfs should be relatively common in the Galaxy, and they
dominate the gravitational wave foreground in the milli-Hertz frequency range \citep{nissanke12,korol17}. Radial velocity surveys
targeting low-mass white dwarfs \citep{marsh95,kilic10a,brown10,brown20} and high-cadence, wide-field photometric
surveys \citep{burdge19a,burdge19b,burdge20} have been successful in finding short period
double white dwarfs. However, low-mass white dwarfs typically outshine their companions, and SB2 systems have
been elusive. \citet{saffer88} identified L 870-2 as the first SB2 white dwarf binary with a period of 1.6 d. However, in the following
three decades, only about two dozen additional systems have been identified \citep{napiwotzki20}. 

Trigonometric parallax measurements provide an opportunity to find SB2 white dwarfs through their over-luminosity.
\citet{bedard17} used a sample of 219 white dwarfs with parallax measurements to identify more than a dozen
over-luminous white dwarfs, and \citet{kilic20} confirmed binarity in at least nine out of 13 of these systems, including four SB2
white dwarfs. Similarly, \citet{hollands18} analyzed the nearly complete Gaia 20 pc white dwarf sample of 139 stars,
and identified several over-luminous binary candidates, including L 870-2.

Here we present high-resolution spectroscopy of two new SB2 white dwarfs, WD 1534+503 and PG 1632+177, and constrain
their orbits. We describe the details of our target selection and observations in Sections 2 and 3, and present the radial velocity
measurements, orbital, and physical parameters of these binary systems in Sections 4, 5, and 6, respectively. We present
a new set of NLTE synthetic spectra for cool white dwarfs along with a comparison with the observed line profiles
in WD 1534+503 and PG 1632+177 in Section 7. We discuss the properties of the current population of SB2 white dwarfs
in Section 8, and conclude.

\begin{figure*}
\centering
\includegraphics[width=2.5in]{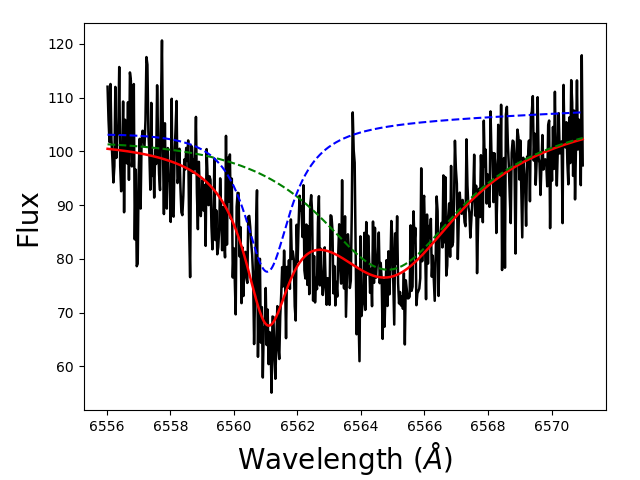}
\includegraphics[width=2.5in]{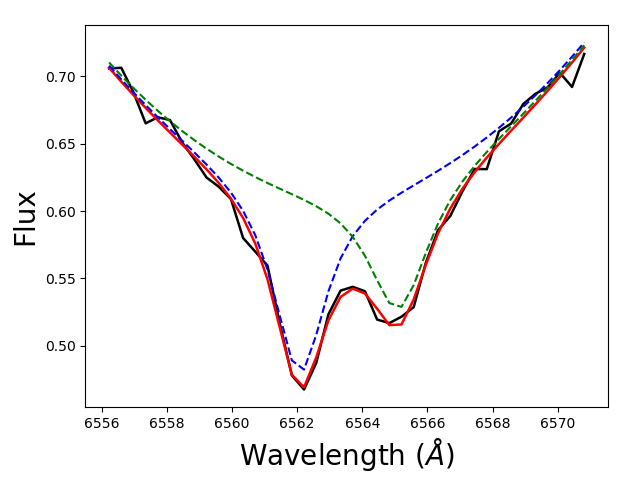}
\includegraphics[width=2.5in]{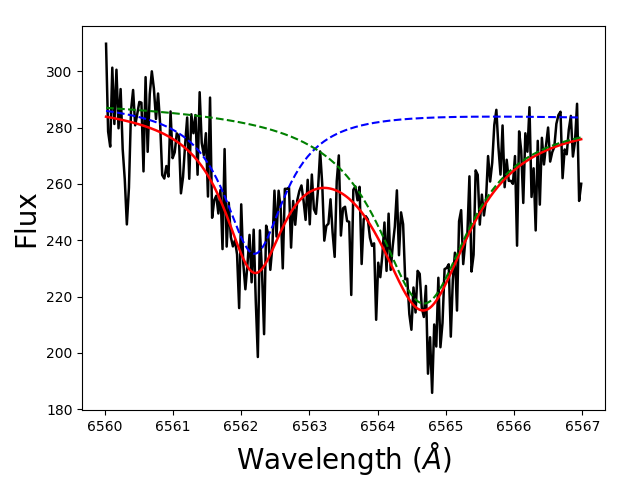}
\includegraphics[width=2.5in]{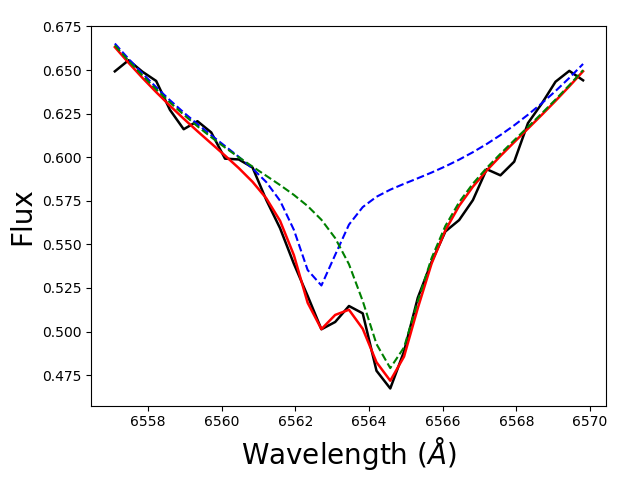}
\caption{Best-fitting Lorentzian profiles (blue and green dotted lines) to the H$\alpha$ line cores visible in the Keck
(left) and Gemini (right) spectra of WD 1534+503 (top panels) and PG 1632+177 (bottom panels). The red solid
lines show the composite best-fitting Lorentzian profiles.}
\label{fitpy}
\end{figure*}

\section{Target Selection}

We selected WD 1534+503 and PG 1632+177 for follow-up observations due to the inconsistencies between their spectroscopic
distance and parallax. Using the spectroscopic method, \citet{gianninas11} derived $T_{\rm eff} = 9010 \pm 130$ K,
$\log{g} = 8.14 \pm 0.05$ for WD 1534+503 based on 1D model atmospheres \citep[see also][]{tremblay11,kleinman13}.
Including the 3D corrections from \citet{tremblay13}, the best-fit parameters are  $T_{\rm eff} = 8960$ K, $\log{g} = 7.87$,
and a spectroscopic distance of 45.1 pc. However, Gaia Data Release 2 parallax \citep{gaia18} puts WD 1534+503 at 68.5 pc. Hence,
it is significantly brighter than expected for a single white dwarf. Interestingly, \citet{zuckerman03} obtained a high-resolution
spectrum of WD 1534+503 to search for metal lines, and they noted the detection of two H$\beta$ components in this system,
and labeled it as a ``possible newly identified double degenerate'' in their Table 2. No further follow-up has been done since then.

Similarly, \citet{gianninas11} used the spectroscopic method to derive $T_{\rm eff} = 10,220 \pm 150$ K, $\log{g} = 8.04 \pm 0.04$
for PG 1632+177. Including the 3D corrections from \citet{tremblay13}, the best fit parameters are $T_{\rm eff} = 10,020$ K,
$\log{g} = 7.80$, and a spectroscopic distance of 17.1 pc. However, Gaia DR2 parallax puts PG 1632+177 at a distance of 25.6 pc,
again indicating that this is also an over-luminous white dwarf. Interestingly, \citet{saffer98} searched
for radial velocity variations in PG 1632+177, but did not find any significant variations. However, their spectral resolution
of 3 \AA\ and their observing cadence of two observations separated by 1-2 h on a single night, followed by a third observation
1 or 2 days later likely made it impossible to detect the double-lines in this $\approx2$ d (see below) orbital period system. 

\section{Observations}

We used the HIRES echelle spectrometer \citep{vogt94} on the Keck I telescope to observe our two targets on UT 2018 June 18.
Due to volcanic activity, our observations were limited to a period of only 2 hours, over which we were able to get a single spectrum
of WD 1534+503, and 4 spectra of PG 1632+177. We used the blue cross disperser with a 1.15 arcsec slit resulting in a spectral
resolution of 37,000. We used {\sc MAKEE} to analyze the HIRES data, and detected double H$\alpha$ lines for both objects.

We obtained follow-up optical spectroscopy of both targets using the 8m Gemini telescope equipped  with the Gemini
Multi-Object Spectrograph (GMOS) as part of the queue program GN-2020A-Q-221. We used the R831 grating and a
0.25$\arcsec$ slit, providing wavelength coverage from 4585 \AA\ to 6930 \AA\ and a resolution of 0.98 \AA.
Each spectrum has a comparison lamp exposure taken within 10 min of the observation time. We used the {\sc IRAF GMOS}
package to reduce these data. 

Our initial observing strategy at Gemini was to obtain 2-3 spectra over 4-5 hours on a single night, and repeat this
sequence on additional nights as the queue schedule permitted. This worked well for WD 1534+503. However, we realized
after the initial observations on PG 1632+177 that its orbital period is much longer than 4-5 hours, and we changed
our observing cadence to a single observation per night for the last four epochs. 

\section{Radial Velocity Measurements}

\begin{figure}
\centering
\includegraphics[width=3in, bb=54 54 758 558]{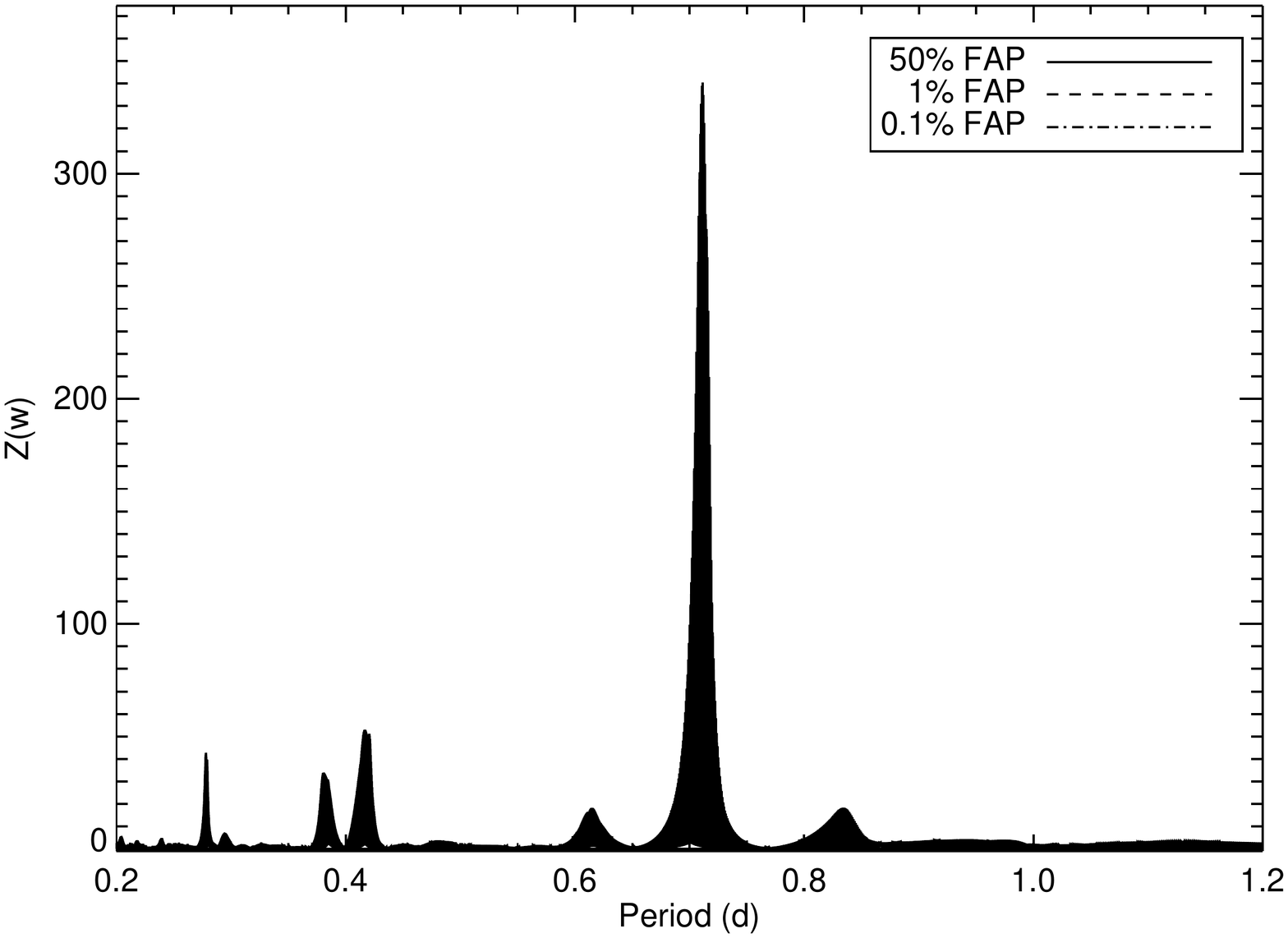}
\includegraphics[width=3in, bb=18 144 592 718]{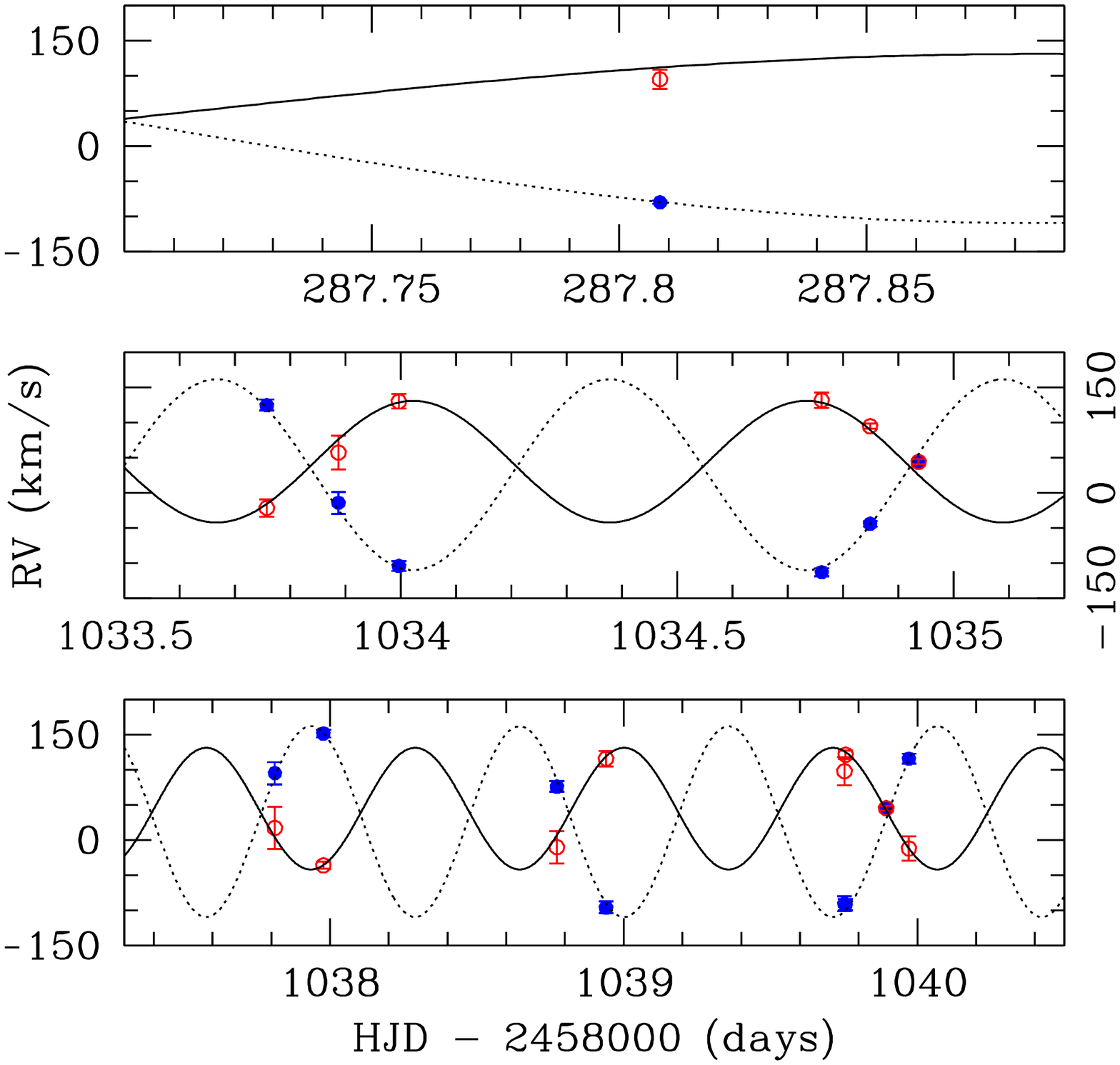}
\caption{{\it Top:} Lomb-Scargle periodogram for WD 1534+503. 
{\it Middle and Bottom:} Radial velocity measurements (open and filled points) of the two components of the
WD 1534+503 system. The solid and dotted lines show the best fit orbit for each component, assuming a circular orbit.}
\label{fig1534}
\end{figure}

\begin{figure}
\centering
\includegraphics[width=3in, bb=54 54 758 558]{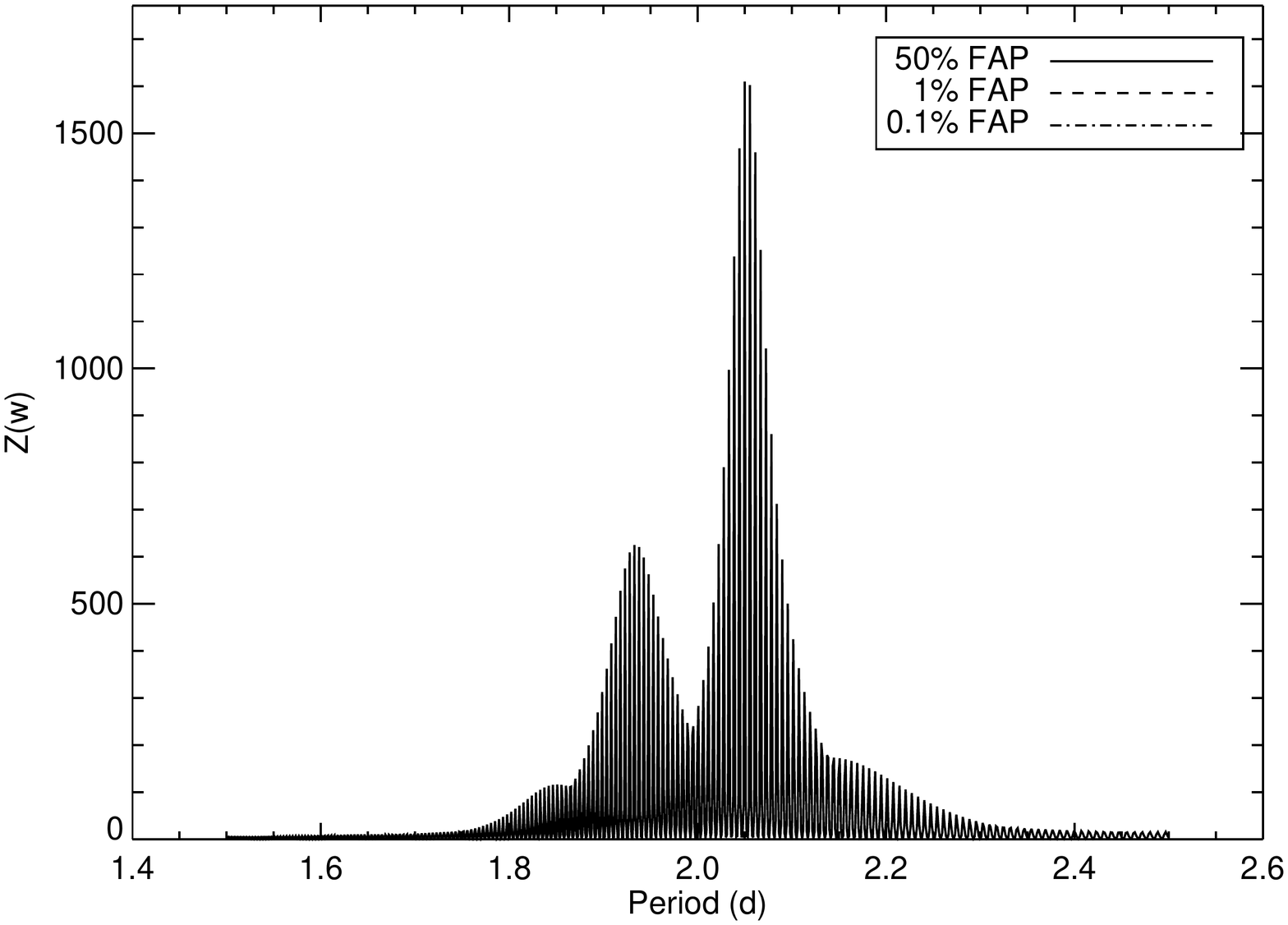}
\includegraphics[width=3in, bb=18 144 592 718]{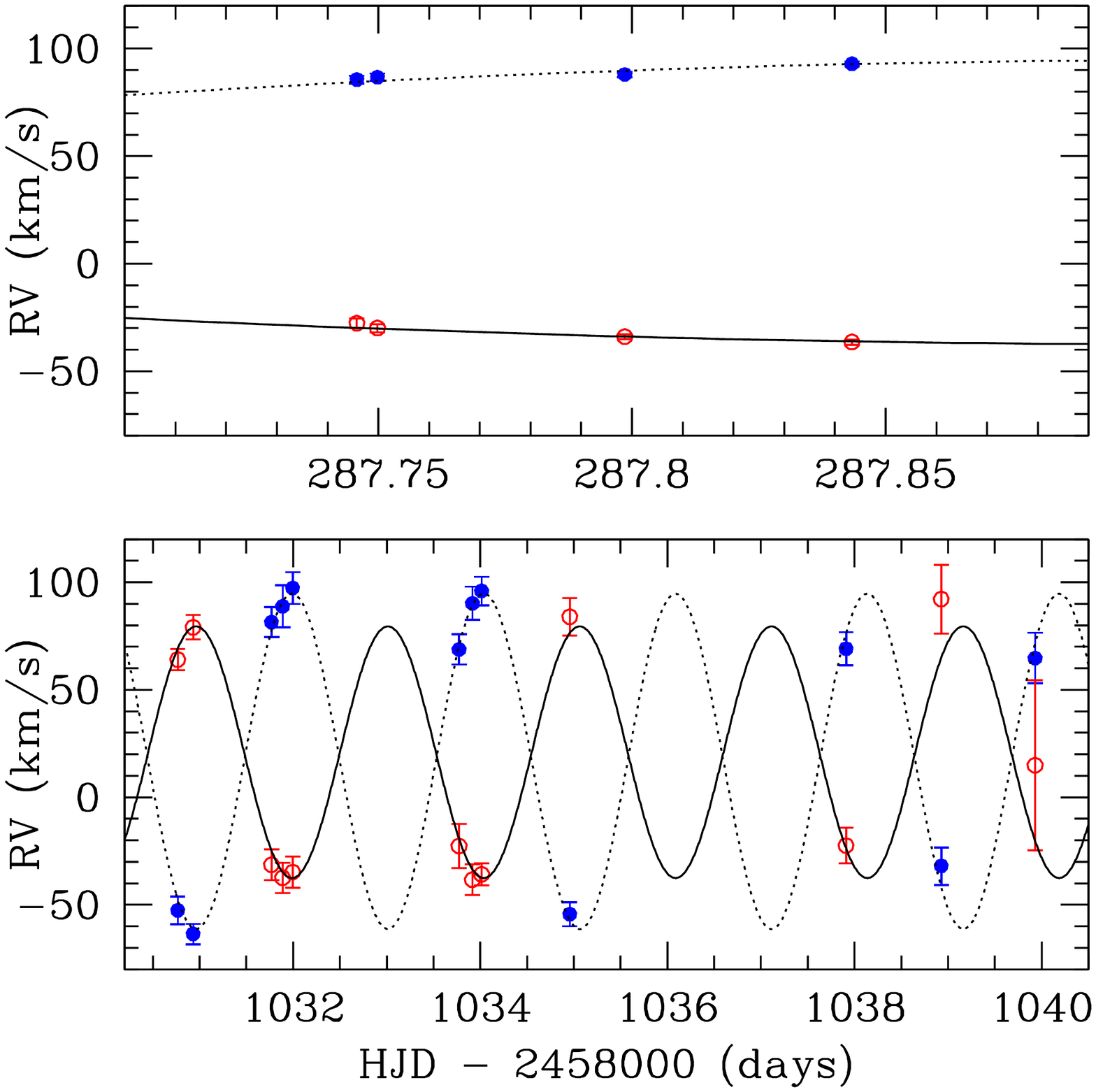}
\caption{{\it Top:} Lomb-Scargle periodogram for PG 1632+177. 
{\it Middle and Bottom:} Radial velocity measurements (open and filled points) of the two components of the
PG 1632+177 system. The solid and dotted lines show the best fit orbit for each component, assuming a circular orbit.}
\label{fig1632}
\end{figure}

We use the same procedures as in \citet{kilic20} to measure radial velocities for our targets.
Briefly, we use a quadratic polynomial plus two Lorentzians (one for each line) to fit the H$\alpha$ line cores.
We use LMFIT, a version of the Levenberg-Marquardt algorithm adapted for Python \citep{newville14}, to find the best-fit
parameters. We apply the standard Solar System barycentric corrections, and use the night skylines to correct
for the spectrograph flexure. 

Figure \ref{fitpy} shows the Keck (left panels) and Gemini (right panels) spectra of WD 1534+503 (top) and PG 1632+177
(bottom) along with the best-fitting Lorentzian profiles to the H$\alpha$ line cores. The red line shows the best-fitting composite
profiles in each case. Here we show Gemini spectra at a similar orbital phase to the Keck spectra so that a fair comparison can be made.
Luckily for both WD 1534+503 and PG 1632+177, one of the H$\alpha$ line cores is significantly deeper than the other, enabling
us to reliably identify the lines at different orbital phases.

We use bootstrapping to estimate the errors in radial velocities as formal fitting errors tend to be underestimated
\citep{napiwotzki20}. We randomly select $N$ points of the observed spectra, where points can be selected more than once,
to rederive velocities, repeating this procedure 1000 times. Tables A1 and A2 present our radial velocity measurements
for WD 1534+503 and PG 1632+177, respectively.  In two of the epochs, we caught WD 1534+503 near conjunction,
with only a single H$\alpha$ line visible in the system. These measurements are included in Table A1, but not used in the orbital
fits as it is impossible to measure the centers for both lines accurately. Similarly, the lines are significantly blended in our last
spectrum of PG 1632+177, and these measurements are included in Table A2, but not used in the orbital fits.

\begin{figure*}
\includegraphics[width=2.5in, angle=270, clip=true, trim=2.2in 0.0in 1.6in 0.0in]{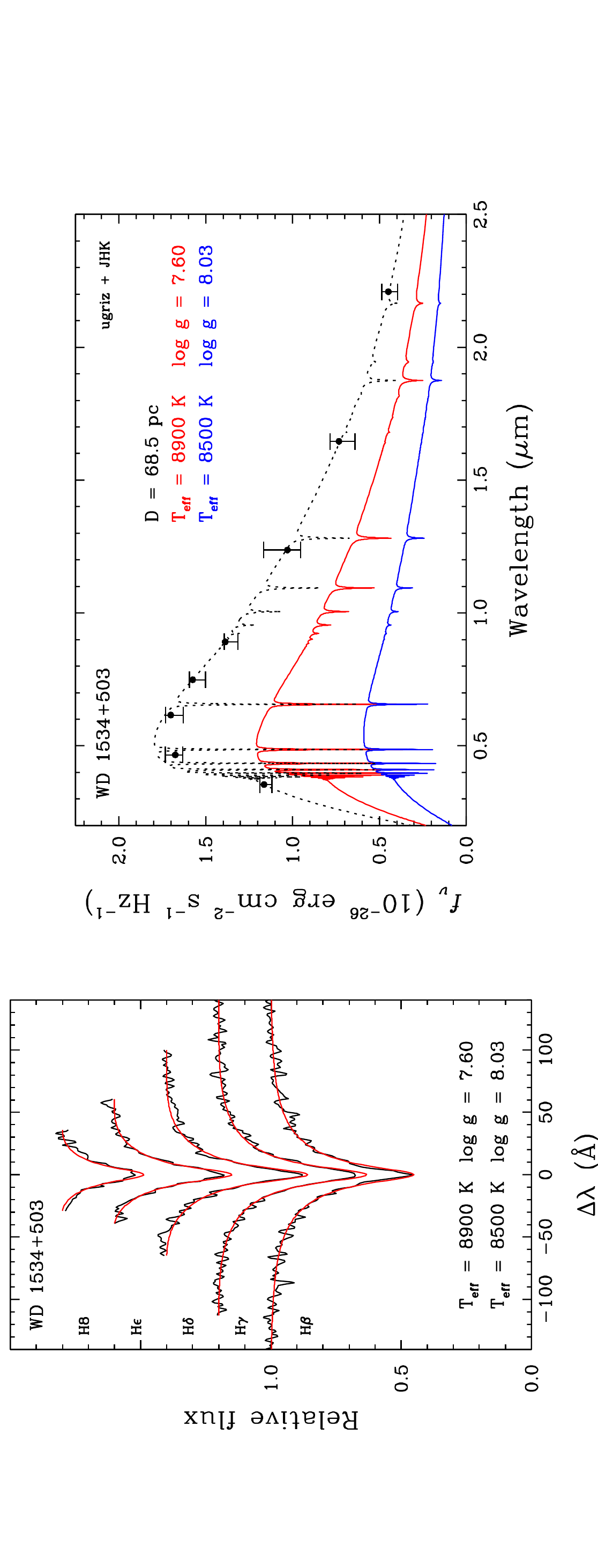}
\includegraphics[width=2.5in, angle=270, clip=true, trim=2.2in 0.0in 1.6in 0.0in]{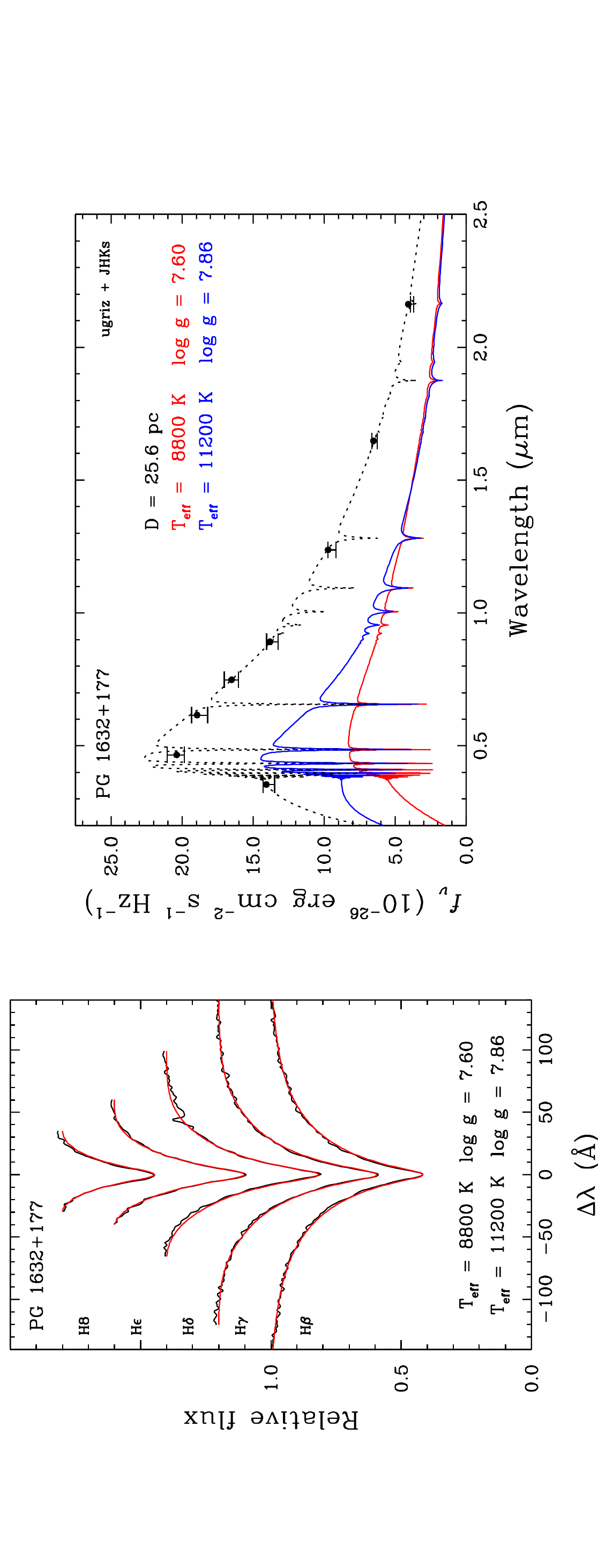}
\caption{Best model-atmosphere fits to the Balmer lines (left panels) and the spectral energy distributions (right panels) of WD
1534+503 and PG 1632+177. In the left panels, the observed and synthetic spectra are displayed as the black and red lines,
respectively. In the right panels, the observed and synthetic average fluxes are shown as the error bars and filled circles,
respectively; in addition, the red and blue lines show the contribution of each component to the total monochromatic model
flux, which is displayed as a black dotted line. The best-fitting atmospheric parameters are given in each panel.}
\label{figsed}
\end{figure*}

\section{Orbital Parameters}

Figure \ref{fig1534} shows the radial velocity measurements and the Lomb-Scargle periodogram for
WD 1534+503. The period is relatively well constrained to 0.71 d in this system. We use the IDL program
MPRVFIT \citep{delee13} in the SB2 mode to find the best-fitting orbit. Excluding the spectra where the
H$\alpha$ lines from both stars overlap and appear as a single line, we have 13 radial velocity measurements.
The solid and dotted lines show the best-fitting orbital solution for each star.

Period aliases are the largest source of uncertainty in the orbital fits. We use a Monte-Carlo approach, re-sampling
the radial velocities with their errors and re-fitting orbital parameters 1000 times. We report the median value and
errors derived from the 15.9\% and 84.1\% percentiles of the distributions for each orbital element. The best-fitting
orbital parameters are $P = 0.71129^{+0.00286}_{-0.00135}$ d, $K_1 = 135.9^{+3.3}_{-3.1}$ \kms,
$K_2 = 86.4 \pm 3.2$ \kms, $\gamma_1 = 25.9^{+2.2}_{-2.1}$ \kms, $\gamma_2 = 45.0 \pm 2.8$ \kms,
$\gamma_2 - \gamma_1 = 19.1 \pm 3.5$ \kms, and ${K_1}/{K_2} =  1.573^{+0.074}_{-0.062}$.

Figure \ref{fig1632} shows the radial velocities and the Lomb-Scargle diagram for PG 1632+177. Excluding a single
spectrum where both lines are blended, we have 15 velocity measurements for this system. The orbital period for this
binary is relatively well constrained to about 2 d, though significant aliasing is present in the Lomb-Scargle diagram.
Performing the orbital fits 1000 times based on a Monte Carlo analysis, the best-fitting orbital elements for PG 1632+177
are $P = 2.04987^{+0.01123}_{-0.00569}$ d, $K_1 = 78.2 \pm 2.0$ \kms, $K_2 = 58.4 \pm 1.9$ \kms,
$\gamma_1 = 16.6 \pm 1.7$ \kms, $\gamma_2 = 20.8 \pm 1.9$ \kms, $\gamma_2 - \gamma_1 = 4.1_{-3.4}^{+2.8}$ \kms, and
${K_1}/{K_2} =  1.342_{-0.056}^{+0.051}$. 

\section{Atmospheric Parameter Determination}

As mentioned in Section 2, the over-luminosity of our targets manifests itself as a severe discrepancy between their
spectroscopic and parallax distances. Another way to look at this is to compare the atmospheric parameters obtained
from spectroscopy and photometry under the assumption of a single star. With this in mind, for each system, we fit
available observed photometry with synthetic photometry computed from single white dwarf model atmospheres (see,
e.g., \citealt{bergeron01}). 

We use SDSS $ugriz$ magnitudes for both targets \citep{ahumada20}, as well as Johnson
JHK magnitudes for WD 1534+503 \citep{zuckerman03} and 2MASS JHK$_{\rm s}$ magnitudes for PG 1632+177 \citep{cutri03}.
We also assume the Gaia DR2 distances \citep{gaia18}. We obtain $T_{\rm eff} = 8870 \pm 260$ K, $\log{g} = 7.26 \pm 0.07$
for WD 1534+503, and $T_{\rm eff} = 10,090 \pm 190$ K, $\log{g} = 7.23 \pm 0.03$ for PG 1632+177. In both cases,
compared to the spectroscopic solutions of \citet{gianninas11} reported in Section 2, the effective temperatures are similar
while the surface gravities are significantly lower. This is typical of unresolved binary systems: a photometric analysis
assuming a single star always yields a very large radius (corresponding to a very low mass white dwarf) to artificially
match the high luminosity produced by the two components (see, e.g., \citealt{bedard17}).

In order to constrain the atmospheric parameters of both components in WD 1534+503 and PG 1632+177, we rely on the
deconvolution procedure introduced by \citet[][see also \citealt{kilic20}]{bedard17}. This method involves fitting simultaneously
the observed Balmer lines and spectral energy distribution with composite model atmospheres. We use the optical spectra
from \citet{gianninas11} that include H$\beta$ through H8, the optical and near-infrared photometry mentioned above, and the
Gaia DR2 parallaxes. The only change in our theoretical framework is that we use the new evolutionary sequences of
\citet{bedard20} in place of the older calculations of \citet{fontaine01}. Note that these sequences are appropriate for
CO-core white dwarfs, while we show below that WD 1534+503 and PG 1632+177 each likely contain a low-mass He-core
component. However, a comparison with the He-core sequences of \citet{althaus13} shows that this small inconsistency
has only a minor impact on our derived parameters (i.e., a change of $\approx$0.03 $M_{\odot}$ for given values of
$T_{\rm eff}$ and $\log{g}$).

A priori, our fitting procedure involves four free parameters: $T_{\rm eff,1}$, $T_{\rm eff,2}$, $\log{g_1}$, and $\log{g_2}$.
However, the individual masses of the components in a white dwarf binary can be derived from the orbital parameters. Since
the difference in systemic velocities is equal to the difference in gravitational redshifts, a combination of this velocity offset
($\gamma_2 - \gamma_1$) with the mass ratio of the binary (derived from ${K_1}/{K_2}$) determines $M_1$ and $M_2$,
and hence $\log{g_1}$ and $\log{g_2}$ given a set of evolutionary sequences. Nevertheless, this approach works well only if
${K_1}/{K_2}$ and ($\gamma_2 - \gamma_1$) are well constrained. For WD 1534+503, there is no significant trend in
${K_1}/{K_2}$ or ($\gamma_2 - \gamma_1$) with the chosen period. However, this is not true for PG 1632+177; there is
a clear trend in the velocity offset based on the best-fit period. For the top four significant aliases between 2.044 and 2.061 d,
${K_1}/{K_2}$ slightly changes from 1.33 to 1.36 with increasing period, but ($\gamma_2 - \gamma_1$) decreases
from $5.5_{-2.8}^{+1.8}$ to $2.5_{-1.7}^{+3.4}$ \kms. Hence, the mass ratio of the binary (through ${K_1}/{K_2}$) is much
better constrained compared to the velocity offset of the two stars. Therefore, for both systems, we rely solely on the mass
ratio in our fitting procedure and use the velocity offset only as a consistency check on our best-fit solution. This means that
$T_{\rm eff,1}$, $T_{\rm eff,2}$, and $\log{g_1}$ are allowed to vary, while $\log{g_2}$ is fixed by the mass ratio.

Figure \ref{figsed} displays our best-fit solutions. Our fitting method yields $T_{\rm eff,1} = 8900 \pm 500$ K, $T_{\rm eff,2} =
8500 \pm 500$ K, $\log{g_1} = 7.60 \pm 0.15$, and $\log{g_2} = 8.03_{-0.16}^{+0.18}$ for the WD 1534+503 system. Both
the spectroscopic and photometric data are nicely reproduced by our composite model. The masses of the two stars are
$M_1 = 0.392_{-0.059}^{+0.069}~M_{\odot}$ and $M_2 = 0.617_{-0.096}^{+0.110}~M_{\odot}$, with an estimated difference
in gravitational redshifts of $16.2_{-4.4}^{+6.3}$ \kms. The latter is entirely consistent with $\gamma_2 - \gamma_1 = 19.1 \pm 3.5$
\kms\ estimated from the radial velocity data.

Similarly, our composite model fit reproduces the spectroscopy and photometry for the PG 1632+177 binary relatively well, with
the best-fit parameters of $T_{\rm eff,1} = 8800 \pm 500$ K, $T_{\rm eff,2} = 11,200 \pm 500$ K, $\log{g_1} = 7.60 \pm 0.15$,
and $\log{g_2} = 7.86_{-0.16}^{+0.17}$.  The masses of the two stars are $M_1 = 0.392_{-0.059}^{+0.069}~M_{\odot}$ and
$M_2 = 0.526_{-0.082}^{+0.095}~M_{\odot}$, with an estimated difference in gravitational redshifts of $8.6_{-2.6}^{+3.6}$ \kms.
The latter is higher than the value obtained from the orbital fits, $\gamma_2 - \gamma_1 = 4.1_{-3.4}^{+2.8}$ \kms, but the
$1\sigma$ confidence intervals overlap.  The orbital and physical parameters of both systems are presented in Table 1.
As mentioned above, our mass estimates likely suffer from a small systematic effect due to our use of CO-core models to analyze
the low-mass components. The use of more realistic He-core models would increase the masses by $\approx 0.03~M_{\odot}$.

\begin{table}
\centering
\caption{Orbital and physical parameters of the two binary systems presented in this paper. Note that
masses are obtained using CO-core models. He-core models result in an increase of $\approx0.03~M_{\odot}$
for the low-mass components.}
\begin{tabular}{lcc}
\hline
Parameter & WD 1534+503 & PG 1632+177 \\
\hline
Period (d)       & $0.71129^{+0.00286}_{-0.00135}$ & $2.04987^{+0.01123}_{-0.00569}$\\
$K_1$ (\kms) & $135.9^{+3.3}_{-3.1}$ & $78.2 \pm 2.0$ \\
$K_2$ (\kms) & $86.4 \pm 3.2$ & $58.4 \pm 1.9$ \\
$\gamma_1$ (\kms) & $25.9^{+2.2}_{-2.1}$ & $16.6 \pm 1.7$ \\
$\gamma_2 - \gamma_1$ (\kms) & $19.1 \pm 3.5$ & $4.1_{-3.4}^{+2.8}$  \\
${K_1}/{K_2}$ & $1.573^{+0.074}_{-0.062}$ & $1.342_{-0.056}^{+0.051}$ \\
$T_{\rm eff,1}$ (K) &  $8900 \pm 500$ & $8800 \pm 500$ \\
$T_{\rm eff,2}$ (K) &  $8500 \pm 500$ & $11,200 \pm 500$ \\
$M_1$ ($M_{\odot}$) & $0.392_{-0.059}^{+0.069}$ & $0.392_{-0.059}^{+0.069}$ \\
$M_2$ ($M_{\odot}$) & $0.617_{-0.096}^{+0.110}$ & $0.526_{-0.082}^{+0.095}$ \\
DR2 Parallax (mas) & 14.5891 $\pm$ 0.0348 & 39.0471 $\pm$ 0.0329 \\
EDR3 Parallax (mas) & 14.6603 $\pm$ 0.0284 & 39.0340 $\pm$ 0.0197\\
\hline
\end{tabular}
\end{table}

\begin{figure}
\includegraphics[width=3.4in, clip=true, trim=0.7in 1.3in 0.7in 1.9in]{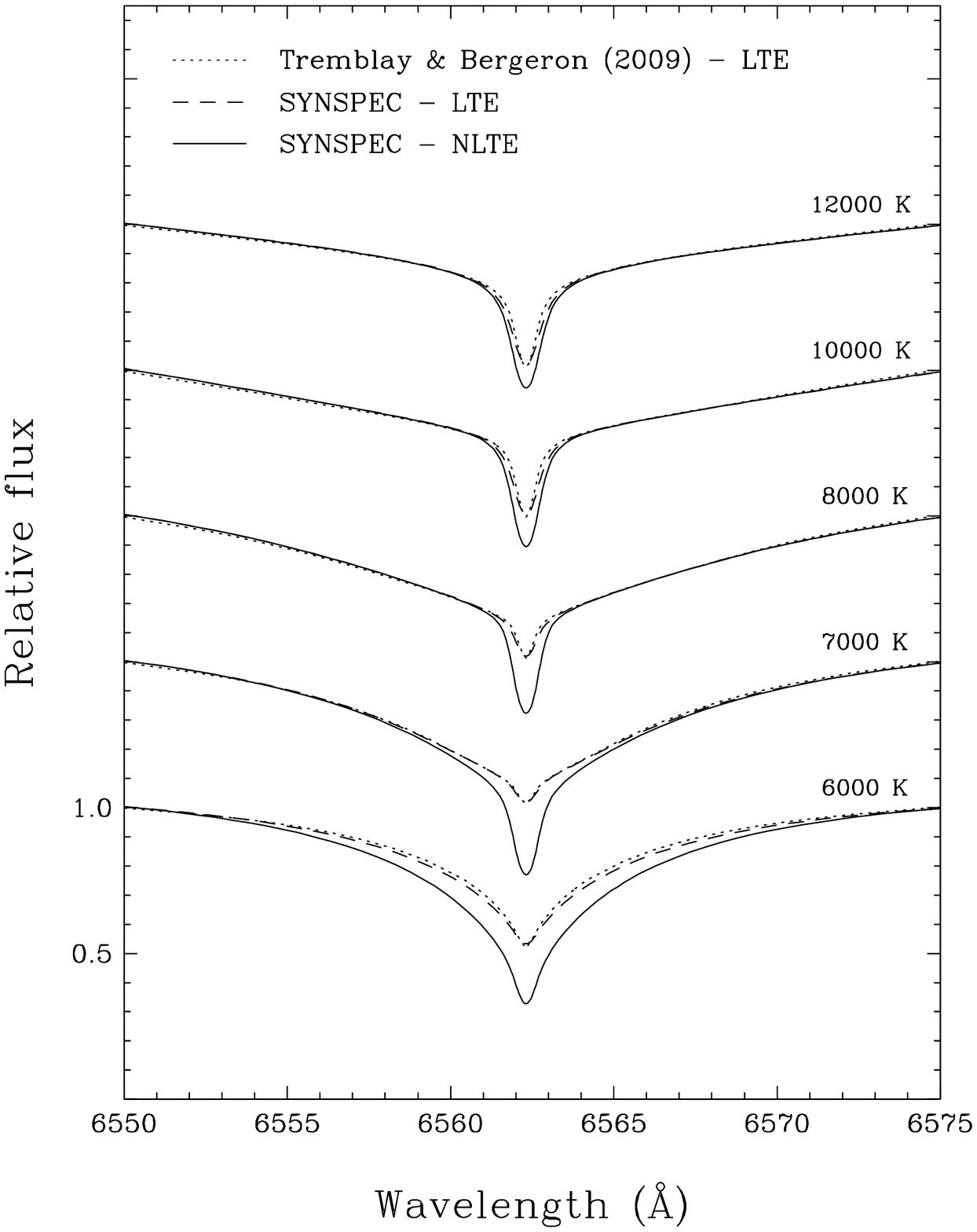}
\caption{Comparison of theoretical H$\alpha$ line profiles at $\log{g} = 8.0$ and various effective temperatures (indicated in
the figure) from three different model grids: the LTE grid of \citet[][dotted curves]{tremblay09}, and our own LTE (dashed
curves) and NLTE (solid curves) grids computed with SYNSPEC. The synthetic spectra are normalized to a continuum
set to unity and are offset vertically by 0.5 for clarity.}
\label{fignlte}
\end{figure}

\begin{figure}
\includegraphics[width=3.2in, clip=true, trim=1.7in 4.2in 1.7in 4.5in]{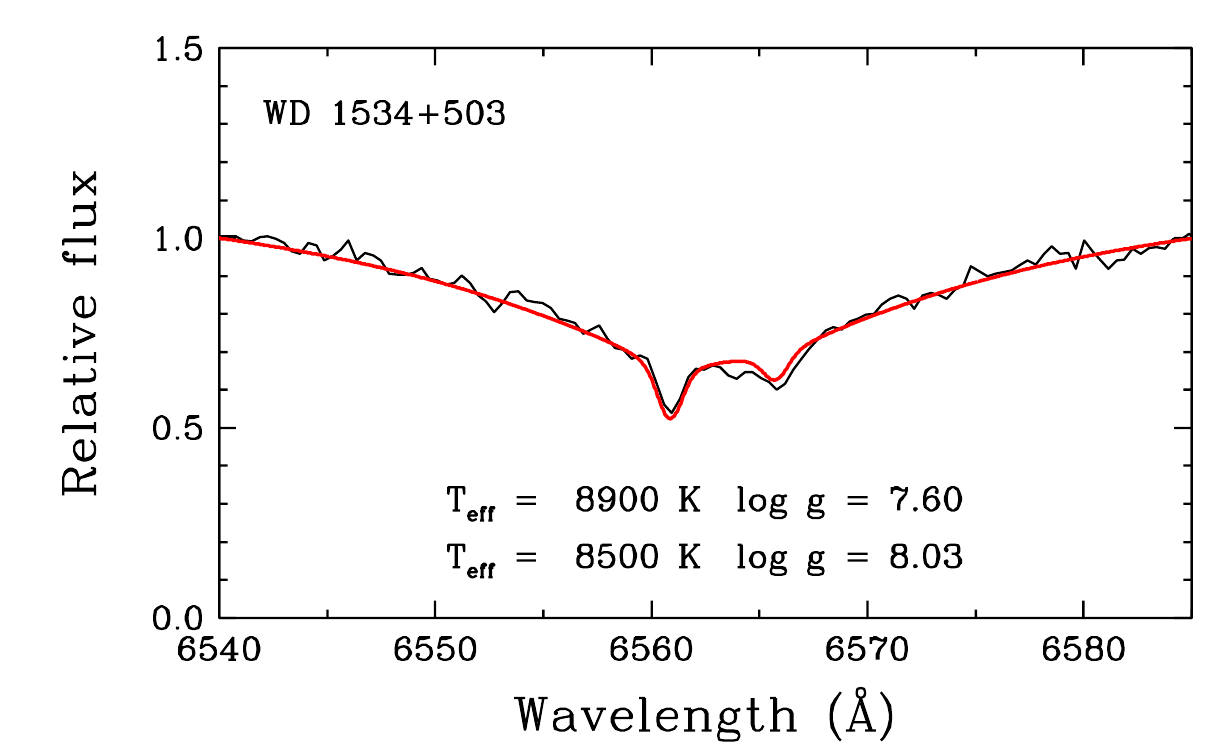}
\includegraphics[width=3.2in, clip=true, trim=1.7in 4.2in 1.7in 4.5in]{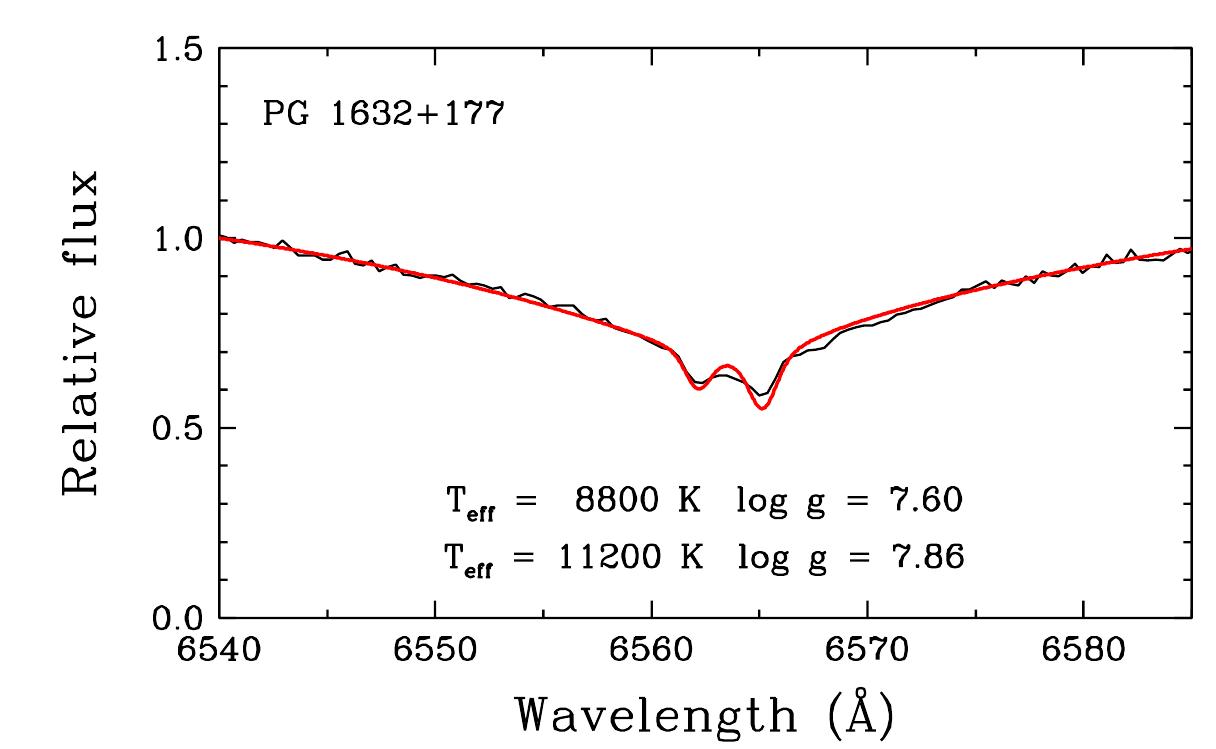}
\caption{Comparison of the observed (black) and predicted (red) double H$\alpha$ features
of WD 1534+503 and PG 1632+177.}
\label{figalpha}
\end{figure}

\section{NLTE Effects in Cool White Dwarfs}

As a further check on our atmospheric parameter determination, we can compare the observed double H$\alpha$ feature
to that predicted by our best-fit solution, as was done by \citet{kilic20} for the two double-lined systems in their sample. In
this comparison, Kilic et al. used the synthetic spectra of \citet{tremblay09} assuming local thermodynamic equilibrium (LTE),
which yielded a reasonably good agreement. Applying the same set of LTE model spectra to the double H$\alpha$ feature of
WD 1534+503 and PG 1632+177, we surprisingly obtain a much poorer agreement, the predicted line cores being too
shallow. Varying the atmospheric parameters only makes the situation worse, suggesting that the problem does not lie in
our deconvolution procedure, but rather in the synthetic spectra themselves. Non-LTE (NLTE) effects appear as the most
plausible explanation, since these are expected to be important in the core of the H$\alpha$ line \citep{heber97,koester98}.

To investigate this idea, we compute NLTE synthetic spectra of H-atmosphere white dwarfs using the code SYNSPEC,
version 51 \citep{hubeny11,hubeny17}. We use the LTE model atmospheres of \citet{tremblay09} as input and perform
NLTE line formation calculations keeping the atmospheric structures fixed. This is an excellent approximation for our
purpose, because the core of the H$\alpha$ line is formed high in the atmosphere, where the radiation field is largely
decoupled from the temperature and pressure structures \citep{heber97,koester98}. In order to model the pressure-broadened
Balmer lines of cool white dwarfs properly, both Stark and neutral broadening must be taken into account \citep{bergeron91}.
We rely on the state-of-the-art Stark profiles of \citet{tremblay09} and on our own implementation in SYNSPEC of a
detailed treatment of neutral broadening, including both resonant and non-resonant processes, following \citet{ali65,ali66}
and \citet{lewis67}. Finally, the continuum opacity of H$^-$, which is significant in cool H-atmosphere white dwarfs, is
considered in our calculations as a ``background'' LTE opacity. Our grid of NLTE synthetic spectra covers $T_{\rm eff} =
5000 - 20,000$ K and $\log{g} = 7.0 - 9.0$. We also generate a similar grid in LTE to allow a direct comparison with the
LTE grid of \citet{tremblay09} and thereby validate our modifications to SYNSPEC.

Figure \ref{fignlte} displays our new NLTE theoretical H$\alpha$ line profiles at $\log{g} = 8.0$ and various effective
temperatures (solid curves). Also shown are the results of our corresponding LTE calculations (dashed curves) as well
as those of \citet[][dotted curves]{tremblay09}. The agreement between both sets of LTE line profiles is excellent, giving
us confidence that we have correctly included the appropriate physics in SYNSPEC. Furthermore, the NLTE
treatment results in deeper line cores, as expected \citep{koester98}. Interestingly, the magnitude of the NLTE effects
actually increases with decreasing effective temperature, contrary to what is seen in
very hot white dwarfs \citep{napiwotzki97}. To our knowledge, this is the first time that this behavior of NLTE effects
in cool white dwarfs is reported. This result nicely explains why LTE line profiles were sufficient to reproduce the H$\alpha$
observations in \citet{kilic20} but not in the present work. Indeed, the double-lined systems analyzed by Kilic et al. contain
relatively hot objects with $T_{\rm eff} \sim 12,000-13,000$ K, for which the difference between the LTE and NLTE line
cores is quite small. On the other hand, WD 1534+503 and PG 1632+177 both include cooler components with
$T_{\rm eff} \sim 8000-9000$ K, for which the NLTE effect is more pronounced.

Figure \ref{figalpha} compares the observed double H$\alpha$ features of both WD 1534+503 and PG 1632+177 with
those predicted by our NLTE calculations using the best-fit atmospheric parameters. We improve the signal-to-noise
by co-adding several of our Gemini spectra at the same orbital phase. The predicted NLTE line cores agree
reasonably well with the observed profiles, though the line core for the hotter component in PG 1632+177 is predicted
slightly too deep. This comparison demonstrates the robustness of our atmospheric solutions, as we simply over-plot
the predicted line profiles from our model fits that do not use these data.

\section{Discussion}

\begin{figure}
\centering
\includegraphics[width=3.2in, bb=18 144 592 718]{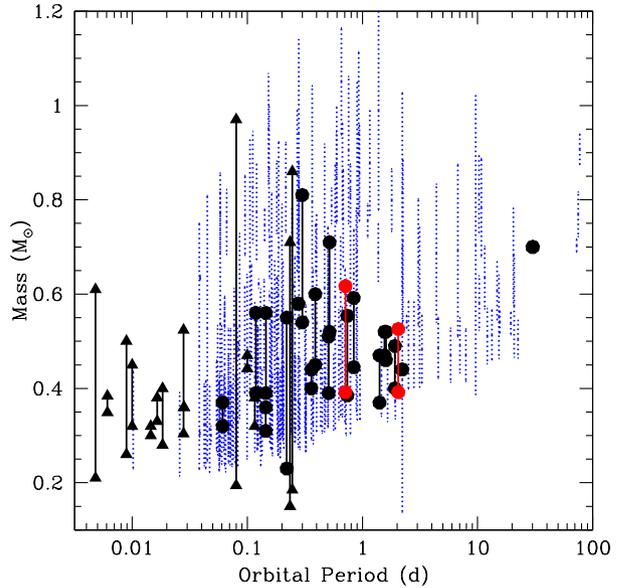}
\caption{Mass and orbital period distribution of known SB2 (circles) and eclipsing (triangles) double white dwarfs
compared to the predictions from binary population synthesis models. The lines connect the components of each
observed (solid lines) and simulated (dotted lines) binary. The red symbols mark WD 1534+503 and PG 1632+177.} 
\label{figdist}
\end{figure}

Figure \ref{figdist} shows the mass and orbital period distribution of all known double-lined spectroscopic binary (SB2) white
dwarfs with orbital constraints, including WD 1534+503 and PG 1632+177, and eclipsing double white dwarfs
\citep[][and references therein]{burdge20,hallakoun16}, along with the predictions from population synthesis models \citep{breivik20}.
The observed population is dominated by low-mass He-core white dwarfs. For example, the two newly discovered systems
presented here, WD 1534+503 and PG 1632+177, both contain low-mass white dwarfs with $M\approx0.39~M_{\odot}$ and
likely CO-core companions. 

There are significant selection biases that favor the discovery of low-mass white dwarf
systems. Since such white dwarfs are significantly larger than their more massive CO core counterparts, they
are over-represented in magnitude-limited surveys, and they are more likely to show photometric effects like eclipses
and ellipsoidal variations, and are therefore easier to discover in transient surveys like the Zwicky Transient
Facility \citep{burdge20}. The shortest period systems, with periods of tens of minutes
\citep{brown11,burdge20}, were found by a highly selective search and cannot be compared to the other white dwarfs
or simulations.

Many SB2 white dwarf binaries are targeted due to their over-luminosity
in color-magnitude diagrams, which again favor nearby, lower-mass systems. Since the detection of the double-lines
typically require high-resolution spectroscopy, the SB2 systems currently known (excluding the eclipsing systems) are
restricted to relatively bright white dwarfs with $G\leq16$ mag.

To simulate the mass and orbital period distribution of short period double white dwarfs, we use the population synthesis
code \textsc{COSMIC} \citep{breivik20} to track the evolution of $10^5$ main-sequence binaries assuming a constant star
formation rate and a 10 Gyr old population. We use independently distributed parameters with primary masses following
the \citet{kroupa93} initial-mass function, a thermal eccentricity distribution, uniformly sampled mass ratios, and
log-uniformly sampled orbital separations, and assume the common-envelope efficiency parameter alpha1 = 1,
and the binding energy factor for common envelope evolution lambdaf = 0.5 \citep[see][for details]{breivik20}. 
We randomly generate a distance to each simulated binary (assuming a constant density) within 100 pc. 

For a fair comparison with the observational sample, here we limit the simulated sample to He- and CO-core white dwarfs, and
only show the simulated binaries brighter than 16th mag, and where both stars in the system have $T_{\rm eff}\geq 6000$ K.
The selection in magnitude ensures that the fainter CO + CO white dwarf binaries are under-represented as in the
observational sample, and the selection in temperature ensures that both white dwarfs would display relatively deep
H$\alpha$ lines, if they have H-rich atmospheres, and therefore these systems would be classified SB2. The dotted lines
in the figure connect the components of each simulated binary. 

Figure \ref{figdist} demonstrates that the orbital period and mass distribution of the observed SB2 and eclipsing
double white dwarfs is remarkably similar to the predictions from the binary population synthesis models. The latter
predict that the lower mass He-core white dwarfs are preferentially found in shorter period systems \citep[see also][]{nelemans01},
which is consistent with the observed sample. The population synthesis models also predict heavier CO + CO white
dwarf binaries at short ($<1$ d) periods, but these tend to be, on average, fainter, and therefore harder to find. Models
also predict binaries with orbital periods longer than a few days. However, the observational sample is significantly
biased against such systems, and currently all but one \citep[WD1115+166,][]{maxted02} of
the SB2 white dwarfs known have orbital periods shorter than about 2.2 d. The identification of longer period systems
is challenging  \citep[see for example][]{napiwotzki20}, but may be possible with large scale astrometric or spectroscopic
surveys like Gaia \citep{andrews19}, the Dark Energy Spectroscopic Instrument (DESI) Milky Way Survey \citep{allende20},
or the SDSS-V \citep{kollmeier19}.

\begin{figure}
\centering
\includegraphics[width=3.2in, bb=20 57 542 744]{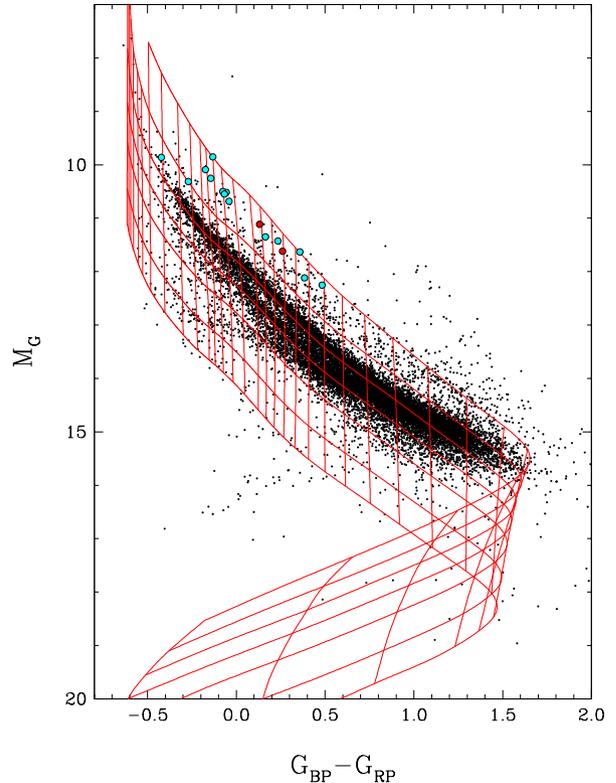}
\caption{Gaia color-magnitude diagram of the 100 pc Montreal White Dwarf Database \citep{dufour17} sample.
Red lines show the cooling sequences for pure-H atmosphere white dwarf models with 0.2, 0.4, 0.6, 0.8, 1.0, 1.2,
and 1.3 $M_{\odot}$ (from top to bottom). Cyan points mark the previously known double-lined spectroscopic binaries
within 100 pc, and the red points mark WD 1534+503 and PG 1632+177.} 
\label{figcmd}
\end{figure}

Figure \ref{figcmd} shows a color-magnitude diagram of the 100 pc white dwarfs from the Montreal
White Dwarf Database \citep[MWDD,][]{dufour17}, along with the cooling sequences for 0.2, 0.4, 0.6, 0.8, 1.0, 1.2,
and 1.3 $M_{\odot}$ pure-H atmosphere white dwarfs. To create a relatively clean white dwarf sample, here we only
include spectroscopically confirmed and candidate (CND) white dwarfs as defined in the MWDD, and exclude the
candidates that appear only in the \citet{gentile19} catalog. The previously known SB2 white dwarfs
and the newly identified systems (WD 1534+503 and PG 1632+177) are marked with cyan and red
symbols, respectively. The current sample of SB2 white dwarfs represents only the tip of the iceberg;
there are a large number of over-luminous white dwarfs within 100 pc of the Sun, $\sim30$\% of which should
be double-lined \citep{kilic20}. Follow-up observations of these over-luminous white dwarfs is guaranteed to
significantly enlarge the SB2 white dwarf population in the solar neighborhood \citep{marsh19}.

\section{Conclusions}

Gaia DR2 parallaxes provide a novel method to identify double white dwarfs through their over-luminosity.
In addition, double-lined systems can be identified based on inconsistencies between their spectroscopic distances
and parallaxes \citep{bedard17}. Here, we present follow-up spectroscopy of two such white dwarfs where the spectroscopic
and parallax distances differ by about 50\%. We show that WD 1534+503 and PG 1632+177 are double-lined
white dwarfs with orbital periods of 0.71 and 2.05 d, respectively. 

We constrain the atmospheric parameters of both components
in each system through a composite model-atmosphere analysis using a new set of NLTE synthetic spectra for
cool white dwarfs. We demonstrate that the NLTE effects in the H$\alpha$ line core increase with decreasing effective temperature.
The predicted NLTE line cores agree well with the observed H$\alpha$ profiles in WD 1534+503 and PG 1632+177.
Both systems contain a low-mass He-core white dwarf with a likely CO-core white dwarf companion.
After L 870-2, PG 1632+177 becomes the second closest double-lined white dwarf binary currently known.

We discuss the orbital period and mass distribution of the SB2 white dwarfs, and
demonstrate that the observed population is consistent with the predictions from the binary population synthesis models,
though the more massive, short period CO + CO white dwarfs are still waiting to be discovered in large numbers.

\section*{Acknowledgements}

We thank Ralf Napiwotzki for a constructive referee report, Jeff Andrews for useful discussions on \textsc{COSMIC},
and Siyi Xu for helping us actively define the cadence of our Gemini observations. 
This work is supported in part by the NSF under grant AST-1906379, the NSERC Canada,
and by the Fund FRQ-NT (Qu\'ebec).

Based on observations obtained at the Gemini Observatory, which is operated by the Association of Universities for Research in Astronomy, Inc., under a cooperative agreement with the NSF on behalf of the Gemini partnership: the National Science Foundation (United States), National Research Council (Canada), CONICYT (Chile), Ministerio de Ciencia, Tecnolog\'{i}a e Innovaci\'{o}n Productiva (Argentina), Minist\'{e}rio da Ci\^{e}ncia, Tecnologia e Inova\c{c}\~{a}o (Brazil), and Korea Astronomy and Space Science Institute (Republic of Korea).

This work was supported by a NASA Keck PI Data Award, administered by the NASA Exoplanet
Science Institute. Data presented herein were obtained at the W. M. Keck Observatory from
telescope time allocated to the National Aeronautics and Space Administration through the
agency's scientific partnership with the California Institute of Technology and the University of
California. The Observatory was made possible by the generous financial support of the W. M.
Keck Foundation.

\section*{Data availability}

The data underlying this article are available in the Gemini Observatory Archive at https://archive.gemini.edu/ and
the Keck Observatory Archive at https://koa.ipac.caltech.edu/cgi-bin/KOA/nph-KOAlogin, and can be accessed with the
program numbers GN-2020A-Q-221 and N018 (or UT 20180618) for Gemini and Keck, respectively.

\input{ms.bbl}

\bsp
\label{lastpage}

\appendix
\section{Radial Velocity Data}

\begin{table}
\centering
\caption{Radial velocities for WD 1534+503}
\begin{tabular}{crr}
\hline
HJD$-$2458000 & $V1_{helio}$ & $V2_{helio}$ \\
(days) & (\kms) & (\kms)\\
\hline
 287.80826950 & $-79.8 \pm 2.4$ & $95.0 \pm 13.7$ \\
1033.75755063 & $125.2 \pm 7.5$ & $-21.2 \pm 12.4$ \\
1033.88721649 & $-14.0 \pm 15.7$ & $57.6 \pm 24.0$ \\
1033.99645563 & $-103.8 \pm 7.0$ & $130.5 \pm 10.3$ \\
1034.76117037 & $-112.8 \pm 5.9$ & $132.0 \pm 11.0$ \\
1034.84874668 & $-43.8 \pm 4.0$ & $95.1 \pm 3.8$ \\
1034.93651786 & $44.3 \pm 3.7$ & $44.3 \pm 3.7$ \\
1037.81231324 & $95.0 \pm 15.6$ & $17.2 \pm 30.1$ \\
1037.97750948 & $151.5 \pm 5.4$ & $-36.0 \pm 4.9$ \\
1038.77259600 & $76.1 \pm 7.8$ & $-10.4 \pm 23.1$ \\
1038.93972014 & $-95.9 \pm 8.2$ & $115.6 \pm 10.9$ \\
1039.75191944 & $-90.9 \pm 10.6$ & $98.0 \pm 20.0$ \\
1039.75559118 & $-92.3 \pm 8.0$ & $121.3 \pm 5.9$ \\
1039.89359319 & $45.3 \pm 3.3$ & $45.3 \pm 3.3$ \\
1039.97106002 & $115.6 \pm 6.9$ & $-12.2 \pm 17.4$\\ 
\hline
\end{tabular}
\end{table}

\begin{table}
\centering
\caption{Radial velocities for PG 1632+177}
\begin{tabular}{crr}
\hline
HJD$-$2458000 & $V1_{helio}$ & $V2_{helio}$ \\
(days) & (\kms) & (\kms)\\
\hline
 287.74574073 & $85.6 \pm 1.8$ & $-27.7 \pm 2.2$ \\
 287.74978969 & $86.8 \pm 1.7$ & $-30.0 \pm 1.9$ \\
 287.79859253 & $87.9 \pm 1.3$ & $-33.9 \pm 1.2$ \\
 287.84337570 & $93.0 \pm 1.2$ & $-36.4 \pm 1.2$ \\
1030.76643626 & $-52.5 \pm 6.6$ & $64.0 \pm 5.0$ \\
1030.93104515 & $-63.6 \pm 4.8$ & $79.1 \pm 5.7$ \\
1031.76860075 & $81.5 \pm 6.9$ & $-31.4 \pm 7.1$ \\
1031.88742845 & $88.8 \pm 9.8$ & $-37.4 \pm 7.1$ \\
1031.99468554 & $97.3 \pm 7.4$ & $-34.8 \pm 7.3$ \\
1033.77338563 & $68.8 \pm 7.1$ & $-22.7 \pm 10.3$ \\
1033.91624887 & $90.2 \pm 7.7$ & $-38.3 \pm 7.2$ \\
1034.01369802 & $95.9 \pm 6.5$ & $-35.8 \pm 5.1$ \\
1034.95654558 & $-54.3 \pm 5.7$ & $83.9 \pm 8.8$ \\
1037.90993764 & $69.1 \pm 7.8$ & $-22.5 \pm 8.3$ \\
1038.92735578 & $-31.9 \pm 8.7$ & $92.2 \pm 15.9$ \\
1039.93019040 & $64.8 \pm 11.8$ & $15.0 \pm 39.5$ \\
\hline
\end{tabular}
\end{table}

\end{document}